\documentclass[journal]{IEEEtran}
\usepackage{cite}
\usepackage{amsmath,amssymb,amsfonts,amsthm,bm}
\usepackage{booktabs,multirow,array}
\usepackage{algorithm,algorithmic}
\usepackage{graphicx}
\usepackage[caption=false,font=footnotesize]{subfig}
\usepackage{xcolor}
\usepackage{microtype}
\usepackage[hidelinks]{hyperref}
\hypersetup{
pdftitle={AdaFCP: Local Neural Functional Representations for Industrial Change Detection},
pdfauthor={Dan Zhuang, Xufei Tang, Yejin Wang, and Houlin Zhou},pdfsubject={},pdfkeywords={}}

\makeatletter
\renewcommand{\section}{\@startsection{section}{1}{\z@}
{2.0ex plus 1.0ex minus 0.5ex}
{0.7ex plus 1ex minus 0ex}{\normalfont\normalsize\centering\scshape}}
\renewcommand{\subsection}{\@startsection{subsection}{2}{\z@}
{2.1ex plus 1.0ex minus 0.5ex}
{0.7ex plus .5ex minus 0ex}{\normalfont\normalsize\itshape}}
\makeatother
\newtheorem{theorem}{Theorem}
\newtheorem{proposition}{Proposition}
\newtheorem{corollary}[proposition]{Corollary}
\newtheorem{remark}{Remark}
\newcommand{\E}{\mathbb E}
\newcommand{\Pbb}{\mathbb P}
\newcommand{\R}{\mathbb R}

\newcommand{\inner}[2]{\left\langle#1,#2\right\rangle}
\newcommand{\argmax}{\operatorname*{arg\,max}}

\begin{document}
\title{AdaFCP: Local Neural Functional Representations for Industrial Change Detection}
\author{Dan Zhuang, Xufei Tang, Yejin Wang, and Houlin Zhou%
\thanks{D. Zhuang is with the School of Mathematics and Statistics, Fujian Normal University, Fuzhou 350117, China.}%
\thanks{X. Tang is with the School of Mathematics and Statistics, Hefei Normal University, Hefei 230601, China.}%
\thanks{Y. Wang is with the School of Big Data and Statistics, Anhui University, Hefei 230601, China.}%
\thanks{H. Zhou (corresponding author) is with the Department of Statistics, Faculty of Arts and Sciences, Beijing Normal University, Zhuhai 519087, China (e-mail: houlin972022@163.com).}}
\maketitle
\begin{abstract}
Industrial sensor measurements often form ordered functional profiles in which persistent operating-regime changes may be confined to short within-cycle regions. Fixed projections can dilute such changes, whereas data-adaptive representation learning complicates reliable statistical testing. This paper proposes adaptive functional change-point detection (AdaFCP) for retrospective detection of a single functional mean change. A compact window and neural basis learn a change-sensitive functional projection by matching the CUSUM path of projected scores to single-change templates. Standardized peak-CUSUM scanning then provides the detection statistic and transition estimate, while the fitted weights identify the contributing within-cycle region. Theoretical analysis characterizes the population path-matching signal and derives an optimal-direction signal-to-noise benchmark. Full-pipeline permutation calibration accounts for adaptive representation learning and restart selection, providing finite-sample false-positive control under exchangeability. A decision-preserving early-stopping variant, AdaFCP-ES, further reduces calibration effort without altering the fixed-budget test decision. Simulations demonstrate advantages over several fixed representations for localized oscillatory changes, while strong analytic and multiscale controls remain competitive. Applications to injection-molding pressure profiles and a nuisance-matched hydraulic benchmark illustrate transition localization and interpretable within-cycle diagnostics, while highlighting the need for dependence-aware calibration in naturally ordered industrial data.
\end{abstract}

\section{Introduction}\label{sec:intro-background}
Industrial monitoring often records complete pressure cycles, vibration traces, or current signatures rather than isolated measurements. Treating these records as ordered functional profiles preserves within-cycle structure and separates two operational questions: \textit{When does the process change across cycles, and which part of a cycle provides evidence of the change}?~\cite{ramsay2005functional,horvath2012inference,wang2016functional} We study retrospective detection of a single persistent mean change, with a sequence-level decision, an estimated onset, and a diagnostic functional direction.

Industrial change-point analysis is relevant to detecting emerging faults and departures from normal operating behavior~\cite{maleki2016changepoint,bilendo2023multitarget}. Functional CUSUM methods provide projection-based and fully functional alternatives~\cite{berkes2009detecting,aue2018detecting}. However, projections based on dominant modes of variation may overlook mean changes occurring in low-variance
directions. Although a sufficiently rich fixed dictionary may contain an informative direction, identifying an effective functional projection from finite observations
remains challenging. This motivates learning projection directions using criteria directly related to the change of interest.

Adaptive functional bases and neural models, including autoencoders, have been used to learn functional representations for prediction and reconstruction~\cite{yao2021deep,thind2023deep,rao2023nonlinear,hsieh2021functional,wu2024autoencoder}.
More directly, kernel-learning and neural approaches to change-point detection use detection-related objectives to learn representations~\cite{chang2019kernel,li2026cusumnet}. These developments suggest the potential of learning functional projections for change-point detection. However, when a projection is learned from the sequence under test, its adaptive selection must be accounted for in statistical calibration. For industrial diagnosis, the learned direction should also be linked to the regions of the profile that contribute to the detection decision, consistent with the emphasis on interpretability in explainable industrial AI~\cite{luo2022xaiindustry,jang2025explainable}.

To address these challenges, we propose adaptive functional change-point detection (AdaFCP), a method that learns a local neural functional direction for detecting changes through CUSUM path matching.
A trainable window selects a within-profile region, and the network learns its contrast. Peak-CUSUM scanning then supplies the decision statistic and location. Because the same sequence trains and tests the representation, every adaptive step is repeated inside permutation calibration. AdaFCP-ES stops when the fixed-budget decision becomes irreversible, without substituting cheaper fits or claiming universal runtime improvement.

The main contributions are summarized as follows.
\begin{enumerate}
\item %\emph{Local Neural Representation:} 
We jointly learn a compact window and neural functional direction through single-change CUSUM path matching. The fitted weights identify within-profile regions contributing to the detection decision.
\item %\emph{Calibration and Theory:} 
We characterize the population path-matching signal and establish finite-sample validity of full-pipeline permutation calibration under exchangeability. Decision-preserving early stopping reduces permutation work without altering the fixed-budget decision.
\item %\emph{Industrial Evaluation:} 
Paired simulations and two industrial-data studies evaluate detection, localization, and within-profile diagnostics. Internal analytic-path and fixed-multiscale controls assess neural parameterization against simpler adaptive searches.
\end{enumerate}

\section{Related Work}\label{sec:related}

\subsection{Functional Representations and Change Detection}

Fourier and spline bases provide standard representations
of functional observations~
\cite{ramsay2005functional,cardot2003spline}.
FPCA extracts leading covariance directions, including
from sparsely observed curves~\cite{yao2005functional}.
Projection-based CUSUM tests operate on functional scores~
\cite{berkes2009detecting}, whereas fully functional
methods avoid initial dimension reduction~
\cite{aue2018detecting}.
Increasing the number of projections can reduce
sensitivity to the chosen projection dimension~
\cite{fremdt2014functional}.
Change-aligned principal components also adapt projections toward potential mean changes~\cite{torgovitski2015changealigned}.
Further developments address multiple functional
change points~\cite{harris2022scalable} and
high-dimensional functional time series~
\cite{li2024highdimensional}.
AdaFCP parameterizes a compactly supported neural functional direction and learns it by matching the projected CUSUM path to single-change templates.

\subsection{Industrial and Neural Monitoring}

Industrial change-point detection is relevant to
emerging-fault monitoring and identifying departures
from normal operating behavior~
\cite{maleki2016changepoint,bilendo2023multitarget}.
Recent developments in deep learning for change-point
detection have been reviewed in~\cite{xu2025deepcpd}.
Representative approaches include neural detectors,
online neural CUSUM, and industrial contrastive
learning~\cite{li2022automaticcpd,gong2022nncusum, liu2024contrastiveindustrial}.

Functional neural models incorporate functional inputs
through basis representations or integration operations~
\cite{thind2023deep,rao2023nonlinear,wang2024functional}.
Adaptive basis networks learn representations for supervised tasks~\cite{yao2021deep}, while functional
autoencoders and shape-informed models address
reconstruction, representation learning, or clustering~
\cite{hsieh2021functional,wu2024autoencoder,singh2025shape}.

Kernel and energy-based methods address more general
distributional changes~
\cite{harchaoui2007retrospective,matteson2014nonparametric},
while recent approaches employ learned kernels or neural critics~
\cite{chang2019kernel,li2026cusumnet}.
CUSUM-Net learns a critic through an aggregate CUSUM objective and applies a CUSUM scan to its one-dimensional output~\cite{li2026cusumnet}. AdaFCP instead fits a compactly supported neural functional direction through single-change template matching. Once fitted, its scalar projection remains linear in the input curve. Full-pipeline calibration repeats direction learning, window fitting, and restart selection for every permutation.
The learned weights provide within-profile diagnostics of the fitted test, rather than causal explanations
or confidence sets for the affected regions.
This distinction is relevant to explainable industrial AI and fault diagnosis~
\cite{adadi2018peeking,barredo2020explainable,luo2022xaiindustry,jang2025explainable}.

\section{Methodology}\label{sec:framework}

The proposed AdaFCP framework consists of local neural
functional projection learning through CUSUM path matching,
peak-CUSUM scanning for detection and localization, and
full-pipeline permutation calibration. An early-stopping
variant, AdaFCP-ES, further reduces calibration effort
while preserving the fixed-budget test decision.

\subsection{Industrial Profiles and Local Projection}\label{sec:model}\label{sec:method}\label{sec:main-adacp}
%Observe scalar curves $X_i(t_j)$, $i=1,\ldots,n$, on $0=t_1<\cdots<t_J=1$, with positive trapezoidal weights $w_j$. 
Observe scalar curves $X_i(t_j)$, $i=1,\ldots,n$,
on $0=t_1<\cdots<t_J=1$, with trapezoidal
quadrature weights $w_1=(t_2-t_1)/2$,
$w_J=(t_J-t_{J-1})/2$, and
$w_j=(t_{j+1}-t_{j-1})/2$ for $2\le j\le J-1$.
The model is
\begin{equation}\label{eq:mean-change-model}
X_i(t)=\mu(t)+\mathbf1\{i>\tau\}\Delta(t)+\varepsilon_i(t).
\end{equation}
The null is $\Delta=0$ and $\tau\in\mathcal K_n=\{\lceil0.1n\rceil,\ldots,\lfloor0.9n\rfloor\}$ under the alternative. Unrestricted permutation requires exchangeability of entire curves.
The evaluated core uses one channel and one projection ($p=d=1$).

To avoid prespecifying the affected region, a trainable window selects where the contrast acts, while a neural function describes its shape. The network acts on the within-curve coordinate. 
With $K_\phi$ Fourier frequencies, its input is
\[
\phi(t)=\bigl(1,t,\{\sqrt2\sin(2\pi rt),\sqrt2\cos(2\pi rt)\}_{r=1}^{K_\phi}\bigr)^\top.
\]
A twice-differentiable scalar-output neural network $f_\theta$ defines
\begin{equation}\label{eq:basis-nn}
\begin{aligned}
g_{u,h}(t)&=[1-\{(t-u)/h\}^2]_+^3,\\
\beta_{\theta,u,h}(t)&=\frac{g_{u,h}(t)f_\theta(\phi(t))}
{\{\sum_jw_jg_{u,h}(t_j)^2f_\theta(\phi(t_j))^2\}^{1/2}}.
\end{aligned}
\end{equation}
Sigmoid parameterizations constrain the center $u$ and halfwidth $h$ to the ranges in Table~\ref{tab:adafcp-configuration}. The twice-differentiable window restricts the learned direction to $(u-h,u+h)$, where the network determines its shape. Weighted normalization ensures unit norm.
Scores are
\begin{equation}\label{eq:score-definition}
C_i=\sum_j w_j\beta_{\theta,u,h}(t_j)X_i(t_j)
\approx\int_0^1\beta_{\theta,u,h}(t)X_i(t)\,dt.
\end{equation}

\subsection{Path-Guided Learning}\label{sec:path-learning}
\begin{figure*}[!t]
	\centering
	\includegraphics[width=0.85\textwidth]{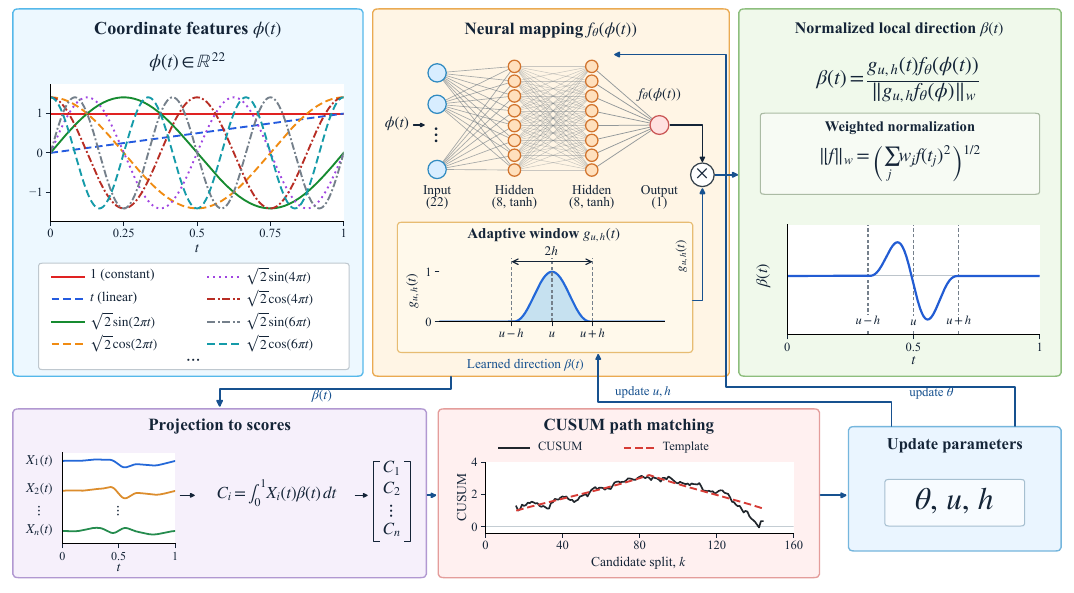}
	\caption{Local neural direction construction and path-guided learning in AdaFCP. Eight of the 22 coordinate features are shown; ellipses indicate omitted features and profiles. A scalar network, adaptive window, and weighted normalization yield $\beta(t)$, and projected CUSUM paths guide regularized updates of $\theta$, $u$, and $h$. Curves are illustrative rather than empirical fits, and the dashed path shows a display-scaled candidate template.}
	\label{fig:local-neural-learning}
\end{figure*}

For a fixed direction, a persistent mean change induces a structured population CUSUM path over candidate splits.
Figure~\ref{fig:local-neural-learning} illustrates how AdaFCP learns a local functional direction by matching the full projected CUSUM path to candidate single-change templates.

Let $m=|\mathcal K_n|$ and let $M\in\R^{m\times n}$ construct $S=MC$ with
\begin{equation}\label{eq:score-cusum}
S_k=\sqrt{\frac{k(n-k)}{n}}
\left(\frac1k\sum_{i\leq k}C_i-\frac1{n-k}\sum_{i>k}C_i\right).
\end{equation}
For every candidate $r$, set $e_r=(\mathbf1\{i>r\})_{i=1}^n$, $v_r=Me_r$, $h_r=v_r/\|v_r\|_2$, and $a_r=m^{-1/2}h_r^\top S$. The templates use candidate positions, not the true change. With $D_i=C_{i+1}-C_i$ and $\bar D=(n-1)^{-1}\sum_iD_i$, use
\begin{equation}\label{eq:score-covariance}
\widehat v=\frac{1}{2(n-1)}\sum_{i=1}^{n-1}(D_i-\bar D)^2,
\qquad\gamma>0.
\end{equation}
The ridge $\gamma$ stabilizes directions with near-zero estimated variance. The same denominator standardizes training and testing. Squared template projections remove sign ambiguity, and log-sum-exp with temperature $\rho>0$ provides a smooth surrogate for selecting a template. Path matching uses
\begin{equation}\label{eq:softmax-cusum}
\mathcal A^{\rm path}_\rho(C)=\rho\log\left[
\frac1m\sum_{r\in\mathcal K_n}
\exp\left\{\frac{a_r^2}{\rho(\widehat v+\gamma)}\right\}\right],
\quad\rho>0.
\end{equation}
For $\Theta=(\theta,u,h)$, minimize
\begin{equation}\label{eq:cp-driven-loss}
-\mathcal A^{\rm path}_\rho(C_\Theta)
+\lambda_{\rm loc}\mathcal R_{\rm loc}
+\lambda_{\rm sm}\mathcal R_{\rm sm}
+\lambda_{\rm null}\mathcal R_{\rm null},
\end{equation}
where
\begin{equation}\label{eq:regularizers}
\mathcal R_{\rm loc}=\sum_jw_j|\beta_\Theta(t_j)|,
\qquad
\mathcal R_{\rm sm}=\sum_jw_j\{\beta_\Theta''(t_j)\}^2,
\end{equation}
\begin{equation}\label{eq:null-stability}
\mathcal R_{\rm null}=L^{-1}\sum_{\ell=1}^L
\mathcal A^{\rm path}_\rho(C_{\pi^{\rm in}_\ell(1)},\ldots,C_{\pi^{\rm in}_\ell(n)}).
\end{equation}
%\textcolor{red}{Under unit weighted norm, $\mathcal R_{\rm loc}$ encourages concentration, while $\mathcal R_{\rm sm}$ discourages rapidly varying weights. The inner-null term penalizes template responses that persist after score reordering, favoring order-sensitive contrasts. Inner reorderings recompute~\eqref{eq:score-covariance}; they regularize learning but do not calibrate the test.} There is no orthogonality penalty for a single direction.
Under unit weighted norm, $\mathcal R_{\rm loc}$ encourages concentration, while $\mathcal R_{\rm sm}$ penalizes curvature. The inner-null term penalizes template responses under reordered scores, favoring
order-sensitive contrasts. Each inner reordering recomputes~\eqref{eq:score-covariance} and regularizes the learning procedure.

To limit dependence on one initialization, each ordering receives $N_{\rm start}$ restarts, each with $N_{\rm upd}$
full-sequence Adam updates. The restart with minimum final penalized loss on $N_{\rm probe}$ prescribed inner
probe permutations is selected. Identical seed schedules, including probes, apply to every outer ordering.
The configuration is summarized in
Table~\ref{tab:adafcp-configuration}.

\subsection{Detection and Full-Pipeline Calibration}\label{sec:score-testing}\label{sec:calibration}
\begin{figure}[!b]
\centering
\includegraphics[width=\columnwidth]{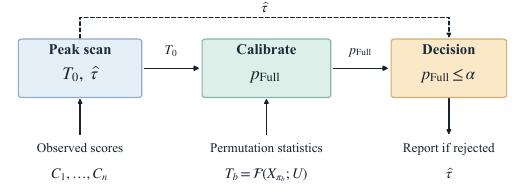}
\caption{Peak scanning and full-pipeline permutation calibration. An onset is reported only after rejection.}
\label{fig:scan-calibration-process}
\end{figure}

Figure~\ref{fig:scan-calibration-process} summarizes the scan and full-pipeline calibration.
After fitting, recompute the scores and define
\begin{equation}\label{eq:dense-stat}
T=\max_{k\in\mathcal K_n}\frac{S_k^2}{\widehat v+\gamma},
\qquad\widehat\tau=\argmax_{k\in\mathcal K_n}\frac{S_k^2}{\widehat v+\gamma}.
\end{equation}
Use the first maximizer for ties. Path matching trains
the representation; its maximizing template is not
the reported location. To account for data-adaptive direction selection, the complete fitting and scanning procedure is repeated for each permutation. 
Conditional on prescribed optimization randomness $U$,
let $\mathcal F(X;U)$ denote the complete fitting and
scanning procedure. Independent uniform permutations
give $T_b=\mathcal F(X_{\pi_b};U)$ and 
\begin{equation}\label{eq:perm-pvalue}
p_{\rm Full}=\frac{1+\sum_{b=1}^B\mathbf1\{T_b\geq T_0\}}{B+1},
\quad T_0=T.
\end{equation}
AdaFCP-Full rejects $H_0$ when $p_{\rm Full}\leq\alpha$
and reports the estimated change-point location only
upon rejection. The fitted direction and its squared
magnitude indicate the within-profile regions emphasized
by the projection, without implying causal attribution.

\subsection{AdaFCP-ES: Decision-Preserving Early Stopping}\label{sec:early-stopping}
Full-pipeline permutation calibration can be computationally expensive. AdaFCP-ES uses bounds on the final permutation $p$-value to stop early when the fixed-budget test decision is guaranteed to remain unchanged.
After $q$ of the fixed $B$ permutations, let $r_q=\sum_{b=1}^q\mathbf1\{T_b\geq T_0\}$. Then
\begin{equation}\label{eq:early-bounds}
L_q=\frac{1+r_q}{B+1}\leq p_{\rm Full}\leq
U_q=\frac{1+r_q+B-q}{B+1}.
\end{equation}
AdaFCP-ES checks these bounds after batches of $G$
permutations. It stops with non-rejection if $L_q>\alpha$
or with rejection if $U_q\leq\alpha$, and otherwise
continues. The procedure reports $[L_q,U_q]$ and
preserves the AdaFCP-Full decision on the same
permutation stream. Computing the exact $p_{\rm Full}$
requires all $B$ permutations. Batch size determines
computational overhead and stopping delay.

\begin{remark}[Asymmetric certificates]
	\label{rem:es-asymmetry}
	For $B=499$ and $\alpha=0.05$, non-rejection requires
	$r_q\geq25$, whereas rejection requires $q-r_q\geq475$.
	Rejection cannot be certified before 475 permutations,
	and batch checks may delay either decision.
\end{remark}

\begin{algorithm}[t]
	\caption{AdaFCP fitting and decision-preserving calibration}
	\label{alg:fit-adabasis}
	\begin{algorithmic}[1]
		\REQUIRE Curves $X$, grid and weights, $\mathcal K_n$,
		seed schedule $U$, $B$, $G$, $\alpha$, and fitting
		settings in Table~\ref{tab:adafcp-configuration}.
		\ENSURE Test decision, calibration result, and
		$\widehat\tau$ upon rejection.
		
		\STATE Define $\textsc{Fit}(Y;U)$ as the complete
		fitting and scanning procedure, including
		$N_{\rm start}$ restarts, $N_{\rm upd}$ Adam updates,
		and selection using $N_{\rm probe}$ inner probes.
		
		\STATE $(\widehat\beta,C,T_0,\widehat\tau)
		\gets \textsc{Fit}(X;U)$.
		\STATE $q\gets0$; $r_q\gets0$.
		
		\WHILE{$q<B$}
		\STATE $g\gets\min(G,B-q)$.
		\FOR{$j=1$ \TO $g$}
		\STATE Draw a uniform permutation $\pi$.
		\STATE $T^\pi\gets
		\textsc{Fit}(X_\pi;U)$
		(test statistic).
		\STATE $r_q\gets r_q+
		\mathbf1\{T^\pi\geq T_0\}$.
		\ENDFOR
		\STATE $q\gets q+g$.
		\STATE Compute $(L_q,U_q)$
		using~\eqref{eq:early-bounds}.
		\IF{ES is enabled and
			($L_q>\alpha$ or $U_q\leq\alpha$)}
		\RETURN Certified decision and
		$[L_q,U_q]$; report $\widehat\tau$
		upon rejection.
		\ENDIF
		\ENDWHILE
		
		\RETURN $p_{\rm Full}=L_B=U_B$
		and its corresponding decision.
	\end{algorithmic}
\end{algorithm}

\subsection{Computation}\label{sec:extensions}

For a fixed projection direction, computing the scores
requires $O(nJ)$ operations, the CUSUM scan requires
$O(n)$, and direct matching against $m$ templates
requires $O(m^2)$. The main computational cost arises
from neural fitting, repeated for up to $B+1$
full-pipeline evaluations. Outer permutations can be
processed in parallel using the prescribed seed schedule.
AdaFCP-ES reduces the number of required outer fits
while preserving the fixed-budget test decision.

\begin{figure*}[!t]
	\centering
	\includegraphics[width=0.82\textwidth,height=0.35\textheight,keepaspectratio]{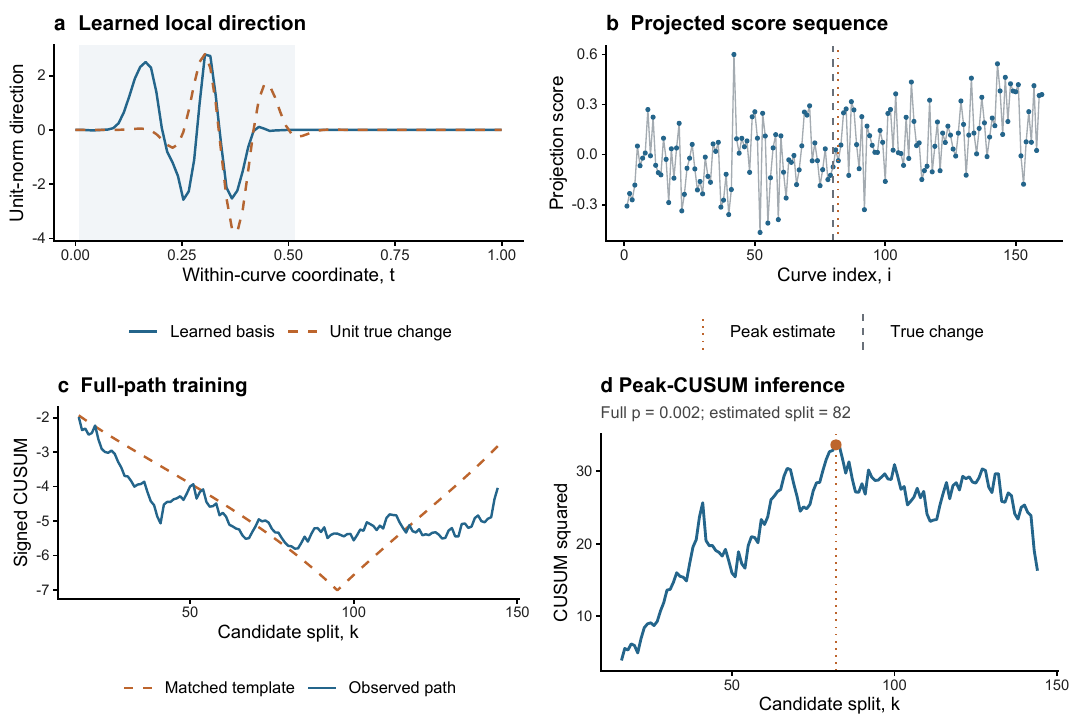}
	\caption{Illustration of AdaFCP for a strong-wavepacket simulation. Panels (a)--(d) show the learned local direction and window, projected scores, CUSUM path matching, and peak-CUSUM detection and localization, respectively. The learned direction is sign-aligned for visualization.}
	\label{fig:learned-mechanism}
\end{figure*}

\section{Statistical Properties}\label{sec:theory}

We characterize the population signal underlying CUSUM
path matching and establish finite-sample validity
of full-pipeline permutation calibration under exchangeability.

\begin{proposition}[Path-matching signal]
	\label{prop:path-signal}
	For a fixed direction $\beta$ with unit weighted norm,
	suppose the errors in~\eqref{eq:mean-change-model}
	have zero mean and a common finite projected variance.
	Let $\delta_\beta=\inner{\beta}{\Delta}$ and
	$v_\beta=\operatorname{Var}(\inner{\beta}{\varepsilon_i})$.
	Then
	\begin{equation}\label{eq:path-signal}
		\begin{aligned}
			\E S&=\delta_\beta v_\tau,\\
			\frac{\{\E a_r\}^2}{v_\beta+\gamma}
			&=\frac{\|v_\tau\|_2^2}{m}
			\frac{\delta_\beta^2}{v_\beta+\gamma}
			\inner{h_r}{h_\tau}^2.
		\end{aligned}
	\end{equation}
\end{proposition}

\begin{proof}
	The mean score vector satisfies
	$\E C=\inner{\beta}{\mu}\mathbf1+
	\delta_\beta e_\tau$.
	Since $M\mathbf1=0$, $\E S=\delta_\beta v_\tau$.
	Consequently,
	$\E a_r=m^{-1/2}\delta_\beta
	\|v_\tau\|_2\inner{h_r}{h_\tau}$,
	which yields~\eqref{eq:path-signal}.
\end{proof}
Equation~\eqref{eq:path-signal} decomposes the squared
population template response into path scale,
regularized projection signal-to-noise ratio (SNR),
and squared template alignment.
The alignment term is bounded by one and attains
its maximum at the true change point.

On the grid, let $W=\operatorname{diag}(w_1,\ldots,w_J)$,
$q=W^{1/2}\Delta$,
$A=W^{1/2}\Sigma W^{1/2}+\gamma I$,
and $z=W^{1/2}\beta$.
The corresponding regularized Rayleigh quotient is
\begin{equation}\label{eq:population-rayleigh}
	Q_\gamma(z)=\frac{(z^\top q)^2}{z^\top Az}.
\end{equation}

\begin{proposition}[Optimal projection benchmark]
	\label{prop:optimal-critic}
	For $A\succ0$ and $q\ne0$, the maximum of
	$Q_\gamma(z)$ over nonzero $z$ is
	$q^\top A^{-1}q$, attained at $z\propto A^{-1}q$.
	For a full-column-rank matrix $V$, the maximum
	over $z\in\operatorname{span}(V)$ is
	$q^\top V(V^\top AV)^{-1}V^\top q$.
\end{proposition}

\begin{proof}
	The unrestricted result follows from Cauchy--Schwarz
	applied to $A^{1/2}z$ and $A^{-1/2}q$.
	Restricting $z=Vc$ and applying the same inequality
	gives the second result.
\end{proof}

The unrestricted optimum provides a benchmark for
the learned local neural direction.

\begin{theorem}[Full-pipeline validity]
	\label{thm:full-permutation}
	Suppose the curves are exchangeable under $H_0$,
	and the prescribed optimization randomness $U$ is
	independent of the data. Let $\mathcal F(\cdot;U)$
	denote the complete fitting and scanning procedure.
	For a fixed $B$, let $\pi_1,\ldots,\pi_B$ be
	independent uniform permutations, independent of
	the data and $U$. Then the permutation $p$-value
	in~\eqref{eq:perm-pvalue} satisfies
	\[
	\Pbb(p_{\rm Full}\leq\alpha)\leq\alpha.
	\]
\end{theorem}

\begin{proof}
	Conditional on $U$ and the permutation orbit,
	the observed and permuted sequences are independent
	and uniformly distributed over the orbit.
	Applying the same measurable procedure yields
	exchangeable test statistics. The plus-one
	permutation $p$-value, with ties counted as
	exceedances, is therefore super-uniform.
\end{proof}

\begin{corollary}[ES decision preservation]
	\label{prop:early-stopping}
	For fixed $B$, $\alpha$, and permutation stream,
	stopping when $L_q>\alpha$ or $U_q\leq\alpha$
	preserves the AdaFCP-Full decision.
\end{corollary}

\begin{proof}
	The final exceedance count lies between $r_q$
	and $r_q+B-q$, yielding
	$p_{\rm Full}\in[L_q,U_q]$.
	Either stopping condition determines the
	fixed-budget decision.
\end{proof}

\section{Experimental Evaluation}\label{sec:simulation}
\subsection{Design and Implementation}
\label{sec:simulation-design}

The main evaluation comprises 3,400 simulated sequences,
with settings determined using independent development data.
We set $n=160$, $J=80$, and $\tau=80$ under the alternatives,
without coordinate-wise standardization.
Table~\ref{tab:adafcp-configuration} summarizes the
prespecified AdaFCP configuration, with fitting settings
held fixed across simulation and industrial applications.
Separate development experiments use $B=199$, with
200 replications per null setting and 100 per
wavepacket alternative.
All baseline methods are evaluated on the same
main simulation sequences.

\begin{table}[t]
\centering\footnotesize
\caption{Prespecified AdaFCP configuration.}
\label{tab:adafcp-configuration}
\setlength{\tabcolsep}{3pt}
\begin{tabular}{@{}p{0.26\columnwidth}p{0.69\columnwidth}@{}}\toprule
Component & Setting\\\midrule
Coordinate input & $K_\phi=10$ Fourier frequencies; 22 inputs\\
Neural map & Two hidden tanh layers, 8 units each; scalar linear output\\
Window search & $u\in[0.10,0.90]$; $h\in[0.08,0.40]$\\
Initialization & $u=0.25,0.50,0.75$; $h=0.25$; $N_{\rm start}=3$\\
Optimization & Adam, learning rate $0.003$; $N_{\rm upd}=40$ full-sequence updates per start\\
Inner reorderings & $L=1$ per update; $N_{\rm probe}=4$ final selection probes\\
Stabilization & Ridge $\gamma=0.01$; temperature $\rho=0.1$\\
Regularization & $\lambda_{\rm loc}=10^{-3}$, $\lambda_{\rm sm}=10^{-7}$, $\lambda_{\rm null}=0.25$\\
Calibration & $B=499$, $\alpha=0.05$; ES batch $G=25$\\\bottomrule
\end{tabular}
\end{table}

Let $\psi_0(t)=1$,
$\psi_{2r-1}(t)=\sqrt{2}\sin(2\pi rt)$, and
$\psi_{2r}(t)=\sqrt{2}\cos(2\pi rt)$ for
$r=1,\ldots,30$. The broad-spectrum null errors are
generated as
\begin{equation}\label{eq:error-expansion}
	\varepsilon_i(t_j)
	=\sum_{\ell=0}^{60}\sqrt{v_\ell}Z_{i\ell}\psi_\ell(t_j)
	+0.15\xi_{ij},
\end{equation}
where $Z_{i\ell}$ and $\xi_{ij}$ are independent
standard normal variables, $f_0=0$,
$f_{2r-1}=f_{2r}=r$, and
$v_\ell=0.035(1+f_\ell/4)^{-0.3}$.
The variances are multiplied by four for
$\ell=0,\ldots,4$.
The hidden-spectrum null retains Fourier frequencies
up to 12, with baseline variances $(\ell+1)^{-1.4}$.
Variance multipliers of 5 and 0.08 are applied to
basis components 2--5 and 10--21, respectively,
using one-based indexing.
Both null settings include the same grid-noise term,
with 500 replications each.

All alternatives use broad-spectrum noise, with
\[
\Delta(t_j)
=\frac{a\,r(t_j)}
{\{\sum_j w_j r(t_j)^2\}^{1/2}},
\qquad a\in\{0.10,0.14\},
\]
and 300 replications per setting.
For each simulated sequence, the applicable
signal parameters are independently generated as
$u_0\sim\operatorname{Unif}(0.25,0.75)$,
$s\sim\operatorname{Unif}(0.05,0.12)$,
$f\sim\operatorname{Unif}(3.5,8.5)$,
$\varphi\sim\operatorname{Unif}(-\pi,\pi)$, and
$\kappa\sim\operatorname{Unif}(-25,25)$.
Let $z=t-u_0$ and $b(t)=\exp\{-z^2/(2s^2)\}$.
The change profiles are defined as
\begin{equation}\label{eq:profiles}
	\begin{aligned}
		r_{\rm GS}(t)
		&=\sqrt{2}\sin(2\pi t),
		&
		r_{\rm LB}(t)
		&=b(t),\\
		r_{\rm WP}(t)
		&=b(t)\cos(2\pi fz+\varphi),\\
		r_{\rm LC}(t)
		&=b(t)\cos\bigl(2\pi[fz+\kappa z^2/2]+\varphi\bigr).
	\end{aligned}
\end{equation}
Global sine (GS) and local bump (LB) serve as
boundary cases, whereas wavepacket (WP) and local
chirp (LC) represent localized oscillatory changes.
The signal-generating parameters are unavailable
to the detection methods.

\subsection{Comparators and Evaluation Criteria}
\label{sec:comparators}

We compare AdaFCP with FPCA-5, Fourier and cubic
B-spline projections using 9, 21, or 41
weighted-orthonormal basis functions, and full-grid
ridge-CUSUM with basis $W^{-1/2}$.
All fixed-representation methods use ridge 0.01.
The order-invariant FPCA basis is cached, while
score variance estimation and CUSUM scanning are
recomputed for each permutation.

We also consider an analytic path control based on
template-wise noise-adjusted projection optimization
and a fixed multiscale control searching 612
windowed directions. The latter combines 17 centers
from 0.1 to 0.9 in increments of 0.05, four
halfwidths (0.10, 0.20, 0.30, and 0.40), and nine
functional shapes consisting of polynomials and
sine/cosine pairs at frequencies 3, 6, and 9.
Both are study-specific structured controls.

Fully functional CUSUM~\cite{aue2018detecting}
is implemented using the R package \texttt{fChange}
(version 0.2.0), with 21 Fourier smoothing functions,
1,000 Brownian-bridge simulations, and lag bandwidth
zero. E-divisive~\cite{matteson2014nonparametric}
uses the R package \texttt{ecp} (version 3.1.5),
with significance level 0.05, at most 499 sequential
permutations per test, minimum segment length 16,
and energy exponent one.
CUSUM-Net~\cite{li2026cusumnet} is implemented
using its publicly available code and the published
aggregate-CUSUM objective. Since its single-change
procedure does not provide a rejection threshold,
it is evaluated only for localization.

Empirical false-positive rates (FPR) and true-positive
rates (TPR) are reported with 95\% Wilson confidence
intervals. Exact McNemar tests based on discordant
paired decisions are used for method comparisons,
with Holm adjustment across a prespecified family
of 90 contrasts. Additional external comparisons
are treated descriptively.

Under the alternatives, unconditional relative
localization error and joint detection--localization
success are defined as
\begin{equation}\label{eq:metrics}
	E=\E\left(\frac{|\widehat\tau-\tau|}{n}\right),
	\qquad
	J=\Pbb\left(
	D=1,\,
	\frac{|\widehat\tau-\tau|}{n}\leq0.05
	\right),
\end{equation}
where $D$ denotes the rejection indicator.
Both measures use all alternative replicates.
E-divisive retains its first accepted interior
split upon rejection and its last considered
split otherwise. Segment-start indices are converted
to preceding-observation indices and restricted to
$1,\ldots,n-1$. Fully functional CUSUM uses its
returned maximizer regardless of the test decision.

The R/torch implementation uses double precision,
one torch/BLAS thread per worker, and two workers.
Runtime includes fitting, calibration, and diagnostics.
Concurrent resource contention is considered when
interpreting computational comparisons.

\subsection{Calibration and Detection Performance}
\label{sec:formal-calibration}

Table~\ref{tab:formal-calibration} summarizes empirical
FPRs under the two null settings.
AdaFCP achieves FPRs of 0.058 and 0.062, with Wilson
95\% confidence intervals of $(0.041,0.082)$ and
$(0.044,0.087)$, respectively, both containing the
nominal level of 0.05.
Fully functional CUSUM and E-divisive also exhibit
FPRs close to the nominal level.
Functional ridge yields a higher FPR of 0.072 under
Broad noise, with a confidence interval of
$(0.052,0.098)$.
The finite-sample validity of AdaFCP follows from
Theorem~\ref{thm:full-permutation} under exchangeability.
AdaFCP-Full requires 81.6 s on average, compared with
3.2 and 9.2 s for the analytic path and fixed multiscale controls, respectively.

\begin{table}[t]
\centering\footnotesize
\caption{Empirical FPRs and mean computation times
	under the two null settings (500 replications, $\alpha=0.05$). External methods use their native calibration procedures; other methods use $B=499$ permutations.}
\label{tab:formal-calibration}
\setlength{\tabcolsep}{3.6pt}
\begin{tabular}{@{}lrrr@{}}\toprule
Method & Hidden & Broad & Time (s)\\\midrule
AdaFCP-Full & 0.058 & 0.062 & 81.6\\
Analytic path & 0.046 & 0.062 & 3.2\\
Fixed multiscale & 0.046 & 0.058 & 9.2\\
FPCA-5 & 0.050 & 0.056 & 0.2\\
Fourier-21 & 0.046 & 0.064 & 0.4\\
B-spline-21 & 0.048 & 0.056 & 0.4\\
Functional ridge & 0.048 & 0.072 & 1.6\\
Fully functional CUSUM & 0.050 & 0.048 & 5.9\\
E-divisive & 0.050 & 0.042 & 0.8\\\bottomrule
\end{tabular}
\end{table}

\begin{figure*}[!t]
\centering
\includegraphics[width=\textwidth,height=0.43\textheight,keepaspectratio]{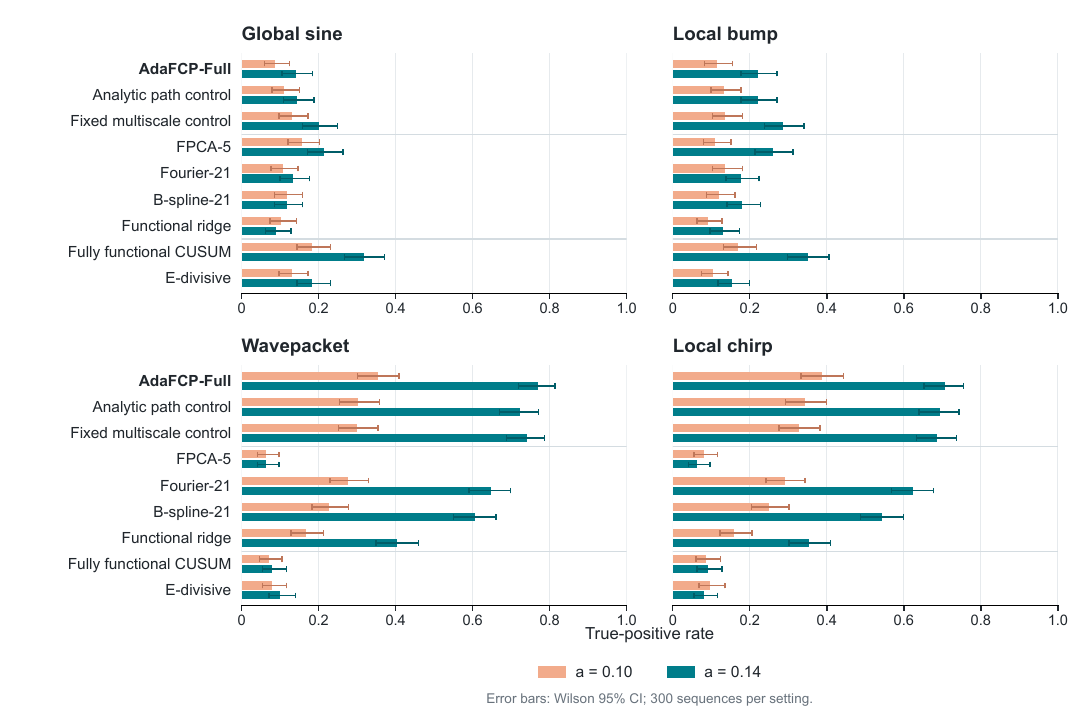}
\caption{TPR over 300 sequences per alternative. Coral and dark teal bars denote $a=0.10$ and $a=0.14$, respectively; horizontal whiskers are marginal Wilson 95\% confidence intervals. Global sine/local bump are boundary regimes, and wavepacket/local chirp are target regimes. Analytic path and fixed multiscale are internal controls. CUSUM-Net has no native single-change rejection rule.}
\label{fig:tpr-all-scenarios}
\end{figure*}

Figure~\ref{fig:tpr-all-scenarios} compares detection
power across the four change scenarios.
For weak and strong wavepackets, AdaFCP achieves TPRs
of 0.353 and 0.770, respectively, while the corresponding
rates for local chirps are 0.387 and 0.707.
For strong wavepackets, the analytic path and fixed
multiscale controls attain TPRs of 0.723 and 0.740.
Compared with B-spline-21, AdaFCP achieves gains of
0.127 and 0.163 for wavepackets and 0.137 and 0.163
for local chirps, all significant after Holm adjustment
($p_{\rm Holm}\leq0.003$).
Against Fourier-21, only the strong-wavepacket gain
of 0.123 remains significant after adjustment
($p_{\rm Holm}=2.77\times10^{-4}$).
Gains over FPCA-5 and functional ridge remain
significant in all four settings, whereas differences
from the analytic path and fixed multiscale controls
do not reach significance.
The fully functional CUSUM and E-divisive procedures
yield TPRs of 0.070--0.100 in these settings.

For global sine changes, AdaFCP achieves TPRs of
0.087 and 0.140, compared with 0.157 and 0.213 for
FPCA-5 and 0.183 and 0.317 for fully functional CUSUM.
For local bumps, AdaFCP attains 0.113 and 0.220,
below the corresponding rates of 0.137 and 0.287
for fixed multiscale and 0.170 and 0.350 for fully
functional CUSUM.
The prespecified primary comparisons in these
boundary settings are not significant after Holm
adjustment, and external comparisons are descriptive.
These results indicate that the advantages of local
adaptation depend on the structure of the change.

\subsection{Localization and Representation Diagnostics}
\label{sec:formal-localization}

Table~\ref{tab:formal-localization} compares unconditional relative localization error $E$ and joint success probability $J$. The analytic path scan achieves the smallest $E$ across all four oscillatory settings, whereas AdaFCP attains the highest $J$ among methods providing a test decision. CUSUM-Net yields localization errors ranging from 0.092 to 0.123 and is excluded from comparisons of $J$ because it does not provide a rejection decision.

\begin{table*}[!t]
\centering\footnotesize
\caption{Oscillatory alternatives: unconditional relative location error $E$ (lower is better) and joint detection--localization probability $J$ (higher is better), using all 300 sequences per cell. Dashes denote unavailable rejection decisions.}
\label{tab:formal-localization}
\setlength{\tabcolsep}{3.8pt}
\begin{tabular}{lrrrrrrrr}
\toprule
& \multicolumn{2}{c}{WP, $a=0.10$} & \multicolumn{2}{c}{WP, $a=0.14$} & \multicolumn{2}{c}{LC, $a=0.10$} & \multicolumn{2}{c}{LC, $a=0.14$}\\
\cmidrule(lr){2-3}\cmidrule(lr){4-5}\cmidrule(lr){6-7}\cmidrule(lr){8-9}
Method & $E$ & $J$ & $E$ & $J$ & $E$ & $J$ & $E$ & $J$\\
\midrule
AdaFCP-Full & 0.100 & \textbf{0.207} & 0.054 & \textbf{0.587} & 0.101 & \textbf{0.233} & 0.056 & \textbf{0.533}\\
Analytic path scan & \textbf{0.083} & 0.173 & \textbf{0.050} & 0.523 & \textbf{0.083} & 0.187 & \textbf{0.053} & 0.480\\
Fixed multiscale scan & 0.132 & 0.157 & 0.067 & 0.540 & 0.132 & 0.187 & 0.067 & 0.510\\
FPCA-5 & 0.232 & 0.010 & 0.237 & 0.013 & 0.235 & 0.020 & 0.237 & 0.003\\
Fourier-21 & 0.143 & 0.133 & 0.080 & 0.440 & 0.142 & 0.133 & 0.088 & 0.380\\
B-spline-21 & 0.146 & 0.110 & 0.091 & 0.387 & 0.147 & 0.127 & 0.093 & 0.340\\
Functional ridge & 0.175 & 0.067 & 0.116 & 0.260 & 0.181 & 0.050 & 0.128 & 0.197\\
\midrule
Fully functional CUSUM & 0.113 & 0.027 & 0.102 & 0.060 & 0.114 & 0.030 & 0.099 & 0.027\\
E-divisive & 0.177 & 0.027 & 0.169 & 0.040 & 0.179 & 0.027 & 0.158 & 0.020\\
CUSUM-Net (public code) & 0.117 & -- & 0.096 & -- & 0.123 & -- & 0.092 & --\\
\bottomrule
\end{tabular}
\end{table*}

Figure~\ref{fig:learned-mechanism} illustrates the fitted representation and detection results for the first strong-wavepacket simulation replicate.
The learned window has center $u=0.261$ and halfwidth
$h=0.253$, capturing part of the localized oscillatory
change. The best-matching template occurs at 95,
whereas the peak-CUSUM estimate is 82, close to the
true change point $\tau=80$, with $p_{\rm Full}=0.002$.

\subsection{Ablation and Capacity Analysis}
\label{sec:formal-ablation}

Component comparisons use the first 200 paired
sequences per null setting and WP amplitude,
separately from the main 300/500-sequence evaluations.
The global/local and path/peak variants share the
same $8\times8$ neural architecture, 40 optimization
updates, and three restarts.
The local linear control has no hidden layer,
whereas QR5 uses a $16\times16$ network, five directions,
and one initialization. These two controls are not
matched in model capacity or computational budget.

\begin{figure}[!tb]
	\centering
	% Stack the existing vector panels; keep the original shared legend below.
	\includegraphics[width=0.82\columnwidth,viewport=0 27 259 232,clip]{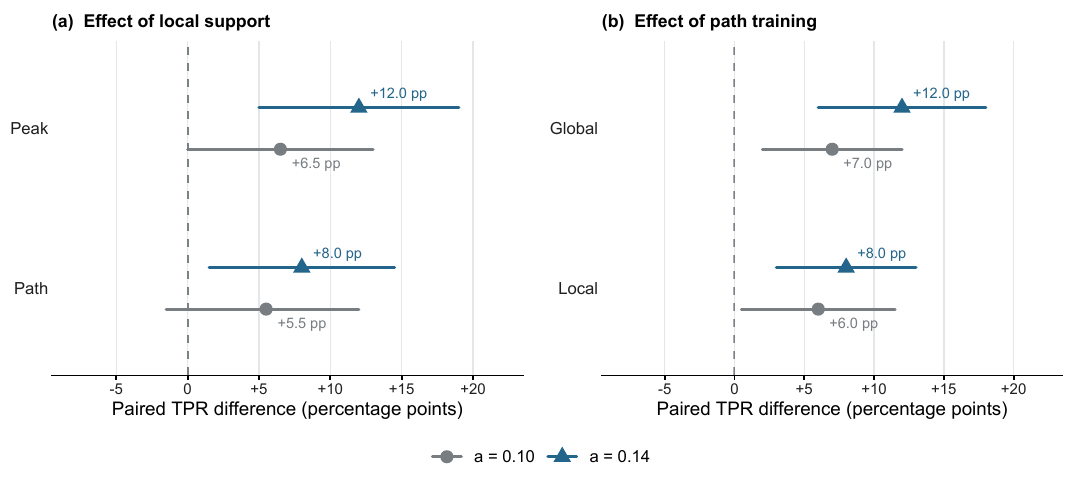}\par
	\smallskip
	\includegraphics[width=0.82\columnwidth,viewport=259 27 518 232,clip]{figures/ablation_paired_effects.pdf}\par
	\includegraphics[width=0.42\columnwidth,viewport=200 3 316 23,clip]{figures/ablation_paired_effects.pdf}
	\caption{Paired TPR differences over 200 WP sequences
		per amplitude: (a) local versus global support;
		(b) path versus peak training.
		Gray and blue denote $a=0.10$ and $a=0.14$,
		respectively. Error bars represent 95\% paired
		bootstrap confidence intervals based on 10,000
		sequence-level resamples.
		Statistical significance is assessed separately
		using Holm-adjusted paired tests.}
	\label{fig:ablation-paired-effects}
\end{figure}

For strong WP, the TPRs are 0.765 (local/path),
0.685 (global/path), 0.685 (local/peak), and
0.565 (global/peak). The corresponding weak-WP
rates are 0.365, 0.310, 0.305, and 0.240.
Figure~\ref{fig:ablation-paired-effects} presents
the paired TPR differences.
Compared with the local linear control, AdaFCP
achieves a gain of 0.135 for strong WP
($p_{\rm Holm}=0.0258$), while the weak-WP gain
of 0.060 is not significant after adjustment.
Both gains over QR5 remain significant
($p_{\rm Holm}<0.001$), whereas differences from
local/peak and global/path do not.
Empirical null FPRs across the six configurations
range from 0.050 to 0.085.
The complete local/path configuration achieves
the highest TPR point estimates, although the
individual contributions of locality and path
matching are not statistically established.

A separate capacity analysis uses 200 development
alternative sequences, with 100 per amplitude.
The fitted local neural directions achieve ridge-SNR
fractions of 0.3895 and 0.5505 relative to the
unrestricted optimum $q^\top A^{-1}q$.
A truth-assisted search with six initializations
and 600 epochs attains fractions of 0.9938 and
0.9932, while the exact Fourier-21 optima attain
0.9870 and 0.9865.
The truth-assisted solutions provide feasible
lower bounds on the maximum achievable neural
ridge-SNR, while the local-support relaxation
provides an upper bound for the corresponding
constrained problem.
The local linear control searches 136 prescribed
windows and additional fitted windows.
The gap between the fitted and truth-assisted
results may reflect estimation error, objective
mismatch, regularization, restart selection,
and optimization.

\subsection{Calibration Cost and Early Stopping}
\label{sec:formal-early-stopping}

AdaFCP-ES preserves the AdaFCP-Full decision for
any fixed permutation stream.
Retrospective replay of all 3,400 Full streams
preserves every decision while reducing the mean
number of evaluated permutations from 499 to 220.9.
Based on measured batch durations, estimated mean
runtime decreases from 81.2 to 37.4 s, corresponding
to a Full/ES ratio of 2.17.
The ratios range from 3.62 to 3.88 under the nulls
and from 1.17 to 1.22 for the strong WP and LC
alternatives. These differences reflect the stopping
asymmetry described in Remark~\ref{rem:es-asymmetry}.

A separate execution study uses 20 saved sequences
from each of the two null and four target settings
($R=120$, $B=499$, $G=25$).
All observed decisions agree with the corresponding
Full results. ES evaluates 265.8 permutations on
average, a reduction of 46.7\%.
The aggregate Full/ES runtime ratio is 1.35,
with ratios of 4.01--4.43 under the nulls and
0.66/0.67 for strong WP/LC.
Because Full reference times and ES execution times
were obtained in different sessions, these ratios
are descriptive and do not provide controlled
wall-clock speedup estimates.
Table~\ref{tab:es-summary} summarizes the replay
and execution results.

\begin{table}[t]
	\centering\footnotesize
\caption{Early-stopping replay and execution results.
	Replay uses 500/300 sequences per null/alternative,
	and execution uses 20 per setting.
	Runtime ratios are Full/ES, with execution references from a separate session.}
	\label{tab:es-summary}
	\setlength{\tabcolsep}{3.6pt}
	\begin{tabular}{@{}lrrrr@{}}
		\toprule
		&\multicolumn{2}{c}{Replay}
		&\multicolumn{2}{c}{Execution}\\
		\cmidrule(lr){2-3}\cmidrule(lr){4-5}
		Setting & Mean $q$ & Ratio & Mean $q$ & Ratio\\
		\midrule
		Hidden null      &114.9&3.88&106.2&4.43\\
		Broad null       &123.0&3.62&128.7&4.01\\
		WP, $a=0.10$     &285.2&1.69&266.0&1.02\\
		WP, $a=0.14$     &423.8&1.17&422.0&0.66\\
		LC, $a=0.10$     &285.8&1.75&273.5&0.98\\
		LC, $a=0.14$     &407.3&1.22&398.3&0.67\\
		\bottomrule
	\end{tabular}
\end{table}

\section{Industrial Applications}
\label{sec:engineering-applications}

\begin{figure*}[!t]
	\centering
	\includegraphics[
	width=0.92\textwidth,
	height=0.22\textheight,
	keepaspectratio
	]{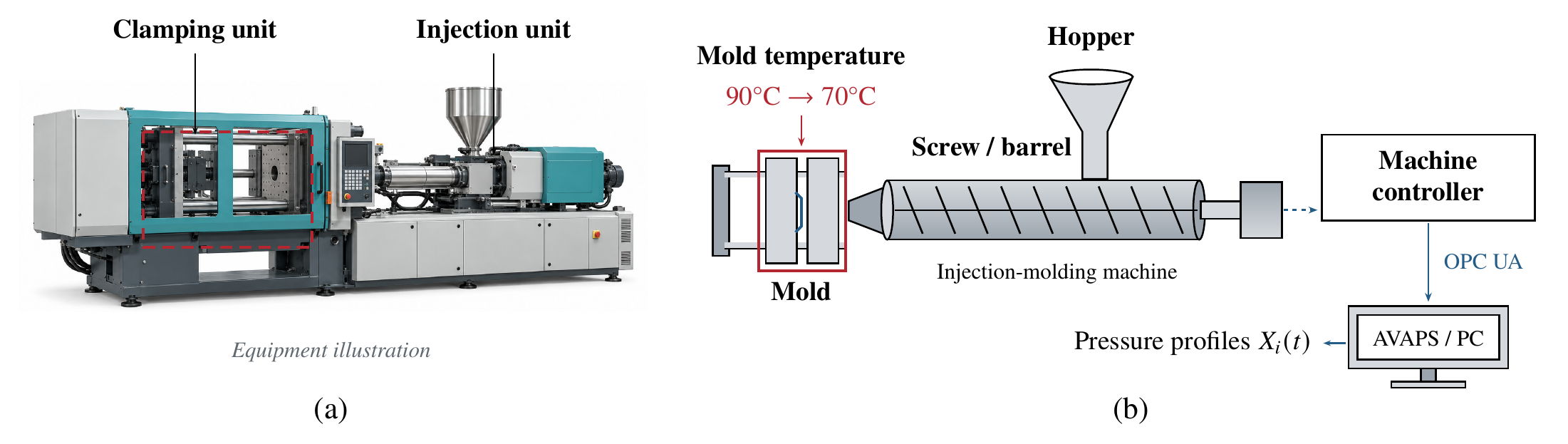}
\caption{Injection-molding process and pressure-data
	acquisition. (a) Illustrative rendering of a generic
	injection-molding machine.
	(b) Schematic of the mold, screw/barrel assembly,
	controller, and OPC UA/AVAPS acquisition system,
	based on the source dataset documentation
	\cite{bogedale2023injection}.
	The illustration does not represent the original
	experimental installation.}
	\label{fig:injection-setup}
\end{figure*}

\begin{figure*}[!t]
	\centering
	\includegraphics[
	width=0.86\textwidth,
	height=0.44\textheight,
	keepaspectratio
	]{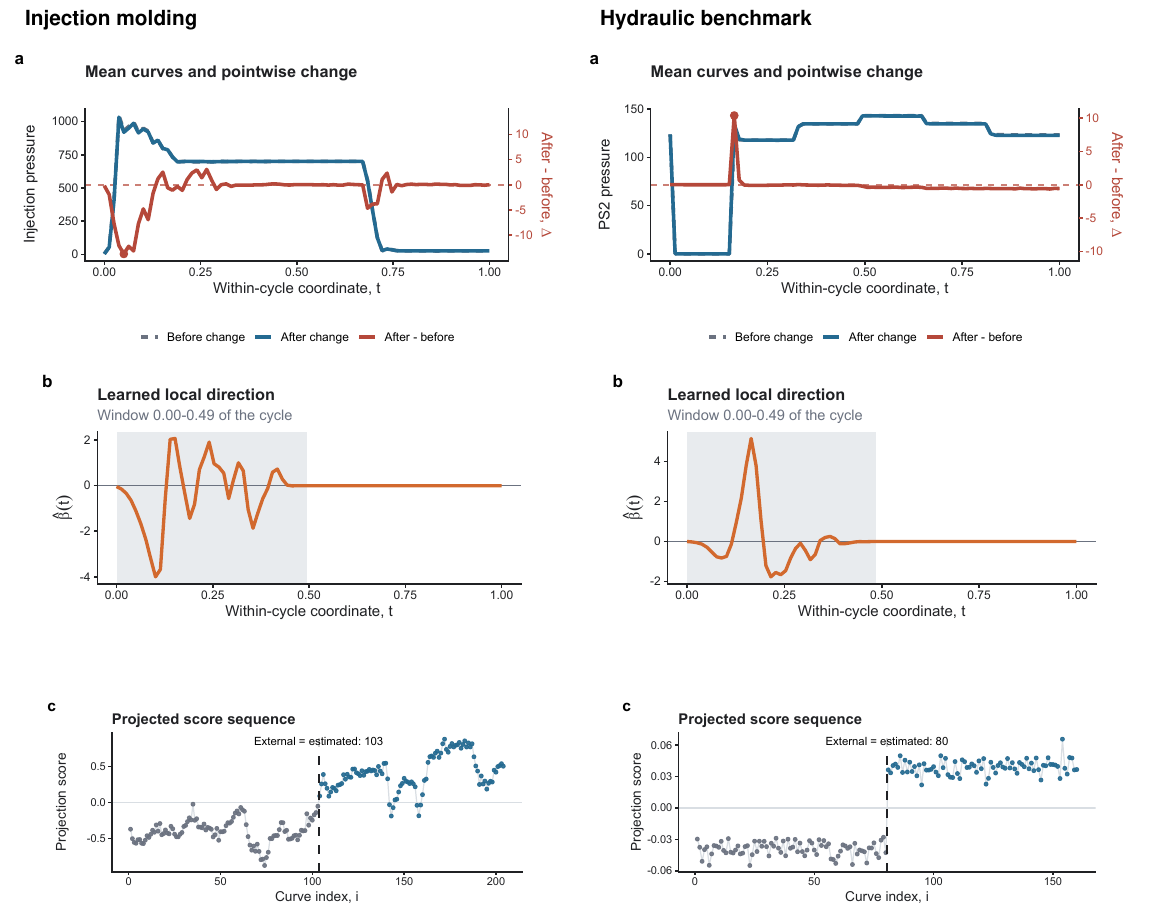}
	\caption{Industrial diagnostics for injection molding
		(left) and matched hydraulics (right).
		Rows display mean profiles and their differences,
		learned directions and windows, and projected scores.
		Projection signs are aligned with the mean differences
		for visualization.}
	\label{fig:applications-compact}
	\label{fig:injection-application}
	\label{fig:hydraulic-application}
\end{figure*}

The application protocols were specified using
public dataset documentation and metadata before
model fitting. Complete curves were downsampled
to $J=80$, centered by the grand mean function,
and scaled by a common pooled RMS, without
coordinate-wise standardization.
AdaFCP and both structured controls retained
their prescribed settings with $B=499$.
Figure~\ref{fig:injection-setup} illustrates the
injection-molding process and pressure-data
acquisition, while Fig.~\ref{fig:applications-compact}
shows the learned representations and measured
profiles.

\subsection{Injection-Molding Process Intervention}

The Stacking Box Data Set I
\cite{bogedale2023injection,scatimdata2023}
contains one injection-pressure curve per production
cycle. HDF5 records are ordered by
\texttt{cycle\_counter}, yielding $n=204$ curves
from cycles 626--829.
The two production blocks, 626--728 and 729--829,
share a reported material moisture content of
0.067\%, while the recorded mold temperature
(\texttt{Twkz}) changes from $90$ to
$70\,^{\circ}\mathrm C$.
This gives an externally documented intervention
point at $\tau=103$.

AdaFCP estimates the change point at 103, with
a learned window of $[0,0.495]$.
The largest-magnitude after-minus-before difference
is $-13.801$ at $t=0.051$, within an early trough
spanning $t=0.03$--$0.12$.
The largest squared basis weights occur over
$t=0.08$--$0.12$, highlighting the early
pressure transient.
The fixed multiscale method also estimates 103,
while the analytic path method estimates 104.
All three nominal permutation $p$-values equal
0.002.

Separately measured mean part weight increases
from 114.299 to 114.488 g, and mean Distance B
increases from 101.548 to 101.707 mm.
These changes provide additional descriptive
evidence of differences between the production blocks.
Within-block lag-one score correlations are
0.869 and 0.923, and both lag-10 Ljung--Box
tests yield $p<0.001$.
Given the substantial serial dependence,
the unrestricted permutation $p$-values are
interpreted descriptively.
The observed intervention alignment and quality
differences do not establish a causal effect.

\subsection{Nuisance-Matched Hydraulic Benchmark}

\begin{figure}[!t]
	\centering
	\includegraphics[
	width=0.96\columnwidth,
	height=0.30\textheight,
	keepaspectratio
	]{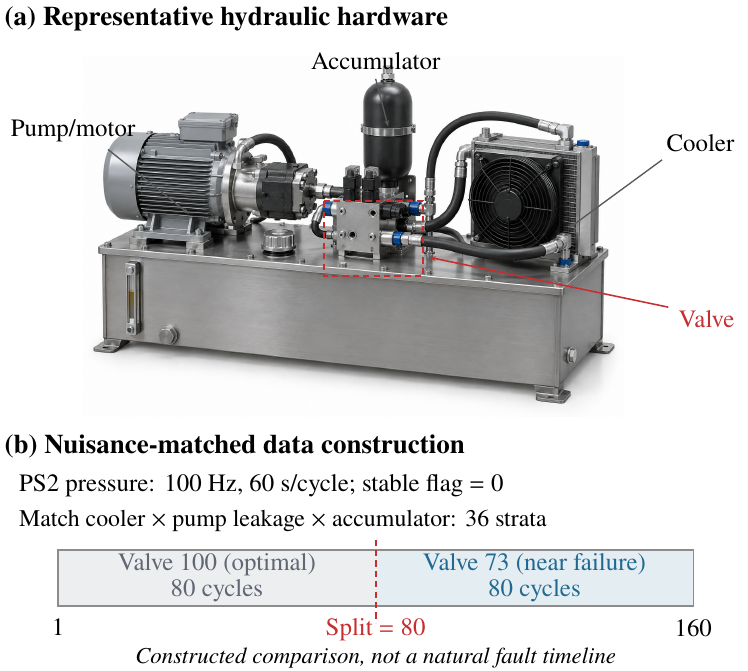}
\caption{Hydraulic benchmark and matched-data
	construction~\cite{helwig2015hydraulic}.
	(a) Illustrative rendering of generic hydraulic
	equipment.
	(b) PS2 pressure curves matched across 36 nuisance
	strata and arranged as 80 cycles per valve state.
	The split at 80 is constructed rather than
	a naturally observed fault onset.}
	\label{fig:hydraulic-setup}
\end{figure}

The UCI hydraulic dataset~\cite{helwig2015hydraulic}
contains 2,205 operating cycles of 60 s each.
Figure~\ref{fig:hydraulic-setup} illustrates the
hydraulic system and benchmark construction.
Because the original condition sequence does not
represent a single change, we construct a matched
benchmark using real PS2 pressure curves.

Cycles with stability flag 0 and valve states
100 and 73 are matched across 36 strata defined
by cooler condition, pump leakage, and accumulator
pressure. Selecting two curves per state from each
stratum and one additional curve per state from
eight prespecified strata yields 80 curves per state.
Sampling and within-state ordering are fixed using
a prespecified random seed.
Valve state 100 precedes state 73, defining a
constructed change point at $\tau=80$.
Outer permutations are restricted within
nuisance strata.

AdaFCP and both structured controls estimate the
constructed change point at 80, with stratified
permutation $p$-values of 0.002.
AdaFCP learns a window of $[0,0.486]$.
The maximum raw pressure increase is 10.389 bar
at $t=0.165$ (9.9 s), while the largest fitted
weights occur over $t=0.14$--$0.18$
(8.4--10.8 s), highlighting a discriminating
pressure transient.
Within-block lag-one score correlations are
0.043 and $-0.195$, with lag-10 Ljung--Box
$p$-values of 0.982 and 0.648, respectively.
The stratified permutation analysis relies on
exchangeability within nuisance strata.
This experiment demonstrates localization of
pressure-profile differences between matched
valve conditions.

\section{Discussion and Conclusion}
\label{sec:discussion}

AdaFCP integrates local neural functional projections,
CUSUM path-guided learning, and full-pipeline permutation
calibration for functional change-point detection.
Simulations demonstrate improved detection and joint
localization performance for localized oscillatory changes
relative to several fixed-representation methods, although
the differences from the analytic path and fixed multiscale
controls are not significant after Holm adjustment.
The advantages are less pronounced for global sine and
local bump changes.
Industrial applications illustrate how the learned
directions identify informative pressure transients,
with comparable change-point estimates obtained by
structured controls.

The current evaluation focuses on single-channel,
single-change settings, with fixed sample size,
observation grid, and change-point location under
independent Gaussian noise in simulations.
Extensions to dependent observations are particularly
important for industrial monitoring, where exchangeability
may not hold.
AdaFCP-ES preserves the fixed-budget test decision
while reducing permutation computations, although
runtime improvements depend on implementation conditions.
Future research will investigate dependence-aware
calibration, more diverse change scenarios, and
computationally efficient representation learning.

\smallskip\noindent\emph{Reproducibility:}
The implementation uses 267 trainable parameters per fit.
Random seeds are fixed for data generation, model fitting,
and permutation calibration.
Reproducibility checks cover data regeneration,
restart and window selection, permutation decisions,
and checkpoint-resume equivalence.
Failed runs are recorded and retained in the evaluation
accounting. The industrial datasets are publicly
available~\cite{scatimdata2023,helwig2015hydraulic}.

\bibliographystyle{IEEEtran}
\bibliography{references}
\end{document}